\documentclass[aps,prc,tightenlines,floatfix,showpacs,showkeys,nofootinbib]{revtex4}

\newcommand{\be}{\begin{equation}}
\newcommand{\ee}{\end{equation}}
\newcommand{\bea}{\begin{eqnarray}}
\newcommand{\eea}{\end{eqnarray}}

\begin{document}

\title{On the possibility to measure $0\nu\beta\beta$-decay nuclear matrix element for 
$^{48}$Ca}


\author{Vadim Rodin}
\address{Institut f\"{u}r Theoretische Physik der Universit\"{a}t T\"{u}bingen, D-72076 T\"{u}bingen, Germany}

\begin{abstract}
As shown in Ref.~\cite{Rod09}, the Fermi part $M_{F}^{0\nu}$  of the total $0\nu\beta\beta$-decay nuclear matrix element $M^{0\nu}$ can be related to the single Fermi transition matrix element 
between the isobaric analog state (IAS) of the ground state of the initial nucleus and the ground state of the final nucleus. The latter matrix element could be measured in charge-exchange reactions. Here we discuss a possibility of such a measurement for $^{48}$Ca and estimate the cross-section of the reaction $^{48}$Ti(n,p)$^{48}$Sc(IAS).

\end{abstract}

\maketitle

Several theoretical approaches have been used to evaluate nuclear matrix elements (NME) $M^{0\nu}$ for neutrinoless double beta ($0\nu\beta\beta$) decay.
There has been great progress in the calculations over the last decade, 
but still there can be a substantial scatter in the calculated $M^{0\nu}$ by different groups. Even more striking, up to a factor of 5, can be the difference in the Fermi part $M^{0\nu}_F$ of the total $M^{0\nu}$.

Therefore, it would be very
important to find  a possibility to determine $M^{0\nu}$ experimentally.
Partial one-leg transition amplitudes to the intermediate $1^+$ states have been measured by charge-exchange reactions in many nuclei (see~\cite{frekers} and references therein), thereby providing important spectroscopic information. 
However, an attempt to reconstruct the nuclear amplitude $M^{2\nu}$ of two-neutrino $\beta\beta$ decay from the measured amplitudes suffers from large inherent uncertainties since relative phases of different intermediate-state contributions cannot be measured. Thus, only if a transition via a single intermediate 
$1^+$ state dominates $M^{2\nu}$, $M^{2\nu}$ can consistently be determined.
Trying the same way to reconstruct $M^{0\nu}$ seems even more hopeless, since  
many intermediate states of different multipolarities (with a rather moderate contribution of the $1^+$ states)
are virtually populated in the $0\nu\beta\beta$ decay due to a large momentum of the exchanged virtual neutrino. In addition, the transition operators in a charge-exchange reaction and $0\nu\beta\beta$ decay become more and more different for higher spins of the intermediate states.

A proposal suggesting a way of a direct measurement of $M^{0\nu}_F$ was put forward in a recent work~\cite{Rod09}. It exploits
the similarity between the Fermi part of the neutrino potential in $0\nu\beta\beta$ decay and the radial dependence of the Coulomb interaction. The latter is well-known to be the leading source of the isospin breaking in nuclei~\cite{auer72
}. As shown in Ref.~\cite{Rod09}, the Fermi matrix element $M_{F}^{0\nu}$ can be related to the single Fermi transition matrix element $\langle 0_f | \hat T^{-} | IAS \rangle$ between the isobaric analog state (IAS) of the  ground state (g.s.) of the initial nucleus and the g.s. of the final nucleus.
Thus, having measured 
$\langle 0_f | \hat T^{-} | IAS \rangle$ by charge-exchange reactions, 
the NME    
$M_{F}^{0\nu}$ can be reconstructed. 

Of course, by measuring only $M_{F}^{0\nu}$ one would not get the total matrix element $M^{0\nu}$ because $M_{F}^{0\nu}\approx -0.3 M^{0\nu}$ ($M^{0\nu}$ is dominated by the Gamow-Teller part $M^{0\nu}_{GT}$). However, the ratio $M_{F}^{0\nu}/M_{GT}^{0\nu}$ may be more 
reliably calculable in different models than $M_{F}^{0\nu}$ and $M_{GT}^{0\nu}$ separately. 
Simple arguments put forward in Ref.~\cite{Rod09} showed that  
an estimate $M_{GT}^{0\nu}/M_{F}^{0\nu}\approx -2.5$ should hold in a realistic calculation (recent QRPA results~\cite{Rod03a} do agree with this simple estimate). 

The master relation, derived in Ref.~\cite{Rod09} in the closure approximation by making use of the isospin symmetry of strong interaction $\hat H_{str}$, represents 
the matrix element $M_{F}^{0\nu}$ in the form of an energy-weighted double Fermi transition matrix element:
\be
M^{0\nu}_F = - \frac{2}{e^2} 
\sum_s \bar\omega_s \langle 0_f | \hat T^{-} |0^+_s \rangle  \langle 0^+_s | \hat T^{-} |0_i\rangle.
\label{MFtot}
\ee
Here, $\hat T^{-}=\sum_{a}\tau_a^{-}$ is the isospin lowering operator,
the sum runs over all $0^+$ states of the intermediate nucleus $_{Z+1}^{\phantom{+2} A} {\mathrm{El}}_{N-1}$, 
$\bar\omega_s=E_s-(E_{0_i}+E_{0_f})/2$
is the excitation energy of the intermediate state $s$ relative to the mean energy of g.s. of the initial and final nucleus.  

As argued in Ref.~\cite{Rod09}, the expression (\ref{MFtot}) in the leading order of the Coulomb mixing must be dominated by the amplitude of the double Fermi transition from the initial g.s. via its IAS into the final g.s.:
\begin{equation}
M^{0\nu}_F \approx - \frac{2}{e^2}\,\bar\omega_{IAS} 
\langle 0_f | \hat T^{-} |IAS \rangle  \langle IAS | \hat T^{-} |0_i\rangle ,
\label{MFappr}
\end{equation}
Here, the second Fermi transition amplitude is non-vanishing due to an admixture of the ideal double IAS (DIAS) wave function
$|DIAS\rangle=\frac{ (\hat T^{-})^2}{\sqrt{4T_0(2T_0-1)}} |0_i^+\rangle$
in the g.s. of the final nucleus: 
$\langle 0_f |\hat T^{-}| IAS\rangle = \langle 0_f | DIAS\rangle \langle DIAS |\hat T^{-}| IAS\rangle$, and $T_0=(N-Z)/2$ is the isospin of the g.s. of the initial nucleus. 

In Eq.~(\ref{MFappr}), the first-leg matrix element $\langle IAS | \hat T^{-} | 0_i \rangle \approx \sqrt{2T_0}=\sqrt{N-Z}$ and the IAS energy $\omega_{IAS}$ are very accurately known.
Thus, the total $M^{0\nu}_F$ can be reconstructed according to Eq.~(\ref{MFappr}), 
if one is able to measure the $\Delta T=2$ isospin-forbidden matrix element 
$\langle IAS | \hat T^{+} | 0_f \rangle$, for instance in charge-exchange reactions of the $(n,p)$-type. 

From the value of $M^{0\nu}_F$ calculated in a model, the magnitude of the matrix element $\langle IAS | \hat T^{+} | 0_f \rangle$ can be estimated by using a transformed version of Eq.~(\ref{MFappr}):
\be 
\langle IAS | \hat T^{+} | 0_f \rangle=  - \frac{e^2 M^{0\nu}_F}{2 \bar\omega_{IAS} \sqrt{N-Z} }.
\label{M}
\ee
Using recent QRPA calculation results for $M^{0\nu}_F$~\cite{Rod03a}, this matrix element can roughly be estimated as $\langle IAS | \hat T^{+} | 0_f \rangle \sim 0.005$,  
i.e. about thousand times smaller than the first-leg matrix element $\langle IAS | \hat T^{-} | 0_i \rangle$. 
This strong suppression of $\langle IAS | \hat T^{+} | 0_f \rangle$ reflects the smallness of the isospin-breaking effects in nuclei. 

The IAS has been observed as a pronounced and extremely narrow resonance, and its various features have well been studied 
by means of $(p,n)$, ($^3$He,$t$) and other charge-exchange reactions on the g.s. of a mother nucleus. In this case the reaction cross-section at the zero scattering angle can be shown to be proportional to a large Fermi matrix element $\langle IAS | \hat T^{-} | 0_i \rangle \approx \sqrt{N-Z}$~\cite{tad87}.
Extraction of a strongly suppressed matrix element $\langle IAS | \hat T^{+} | 0_f \rangle$ from a tiny cross-section of the $(n,p)$-type reactions on the final nucleus might only be possible if there exists a similar proportionality in the $(n,p)$ channel.

As argued in Ref.~\cite{Rod10}, the isospin of the projectile should not be larger than $T=1/2$ ($(n,p)$, ($t,^3$He), \dots reactions). However, it is still not guaranteed that the reaction cross-section $\sigma(0_f^+\to IAS)$ for these probes is proportional to a strongly suppressed matrix element $\langle IAS | \hat T^{+} | 0_f \rangle$, since the other isospin impurities in the wave functions $| 0_f \rangle$ and $| IAS \rangle$ may have a larger effect on the reaction cross-section. 

A preliminary assessment of the $(n,p)$ reaction at the zero scattering angle, performed in Ref.~\cite{Rod10} for an intermediate-mass $\beta\beta$-decaying nucleus $^{82}$Se, has shown that the tiny cross-section $\sigma_{np}(0_f^+\to IAS)$ is indeed dominated by the admixture of the DIAS in the g.s. of the final nucleus, provided that the mixing between the IAS and the $0^+$ states of normal isospin in the intermediate nucleus is weak and can be treated perturbatively.
However, in heavy nuclei the spread of the IAS becomes rather significant and should be taken into consideration. 

A candidate $\beta\beta$-decaying nucleus in which such a perturbative treatment may be justified in reality is $^{48}$Ca. The IAS of $^{48}$Ca is a state with $J^\pi=0^+, T=4, T_z=3$ at the excitation energy of $E_x=$6.678 MeV ($\bar\omega_{IAS}\approx $8.5 MeV) in $^{48}$Sc. It lies under the threshold of particle emission and with almost 100\% probability decays to $1^+$ state at $E_x=$2.517 MeV by the emission of a $\gamma$-quantum with $E_\gamma$=4.160 MeV~\cite{Folk75}. This $\gamma$-decay energy is much higher than the $\gamma$-decay energies from $0^+$ states with normal isospin $T=T_z=3$ surrounding the IAS (which decay by a cascade), and could be used as a unique experimental tag telling that the IAS indeed was excited in a reaction.

There are strong arguments that the IAS of $^{48}$Ca must be a single state without fragmentation. 
The state-of-the-art measurement of $^{48}$Ca($^{3}$He,$t$)$^{48}$Sc(IAS) reaction~\cite{Gr07} does in fact contribute to clarification of this issue as discussed below.

Fragmentation of the IAS may occur only if there are several $0^+$ states with the normal isospin around the IAS to which the IAS may strongly couple. 
In other words, the total number of the $0^+$ states within the IAS spreading width 
$\Gamma^\downarrow_A$ must be greater than one (for nuclei around $A=50$ \ \ $\Gamma^\downarrow_A$  is typically about few keV).
In the back-shifted Fermi-gas model~\cite{Dilg73} the level density $\rho(U,J,\pi)$
of $J^\pi$ states is given by:
\begin{equation}
\rho(U,J,\pi)= {2J+1 \over 48\sqrt{2}}{1 \over  \sigma^3
a^{1/4}}{\exp\left(2\sqrt{aU}-{J(J+1) \over
2\sigma^2}\right) \over (U+t)^{5/4}}
\end{equation}
Here,  $U=E-\delta$ is the effective excitation energy with $\delta$ being the backshift($\delta>0$ for even-even, $\delta\approx 0$ for odd-A, $\delta<0$ for odd-odd nuclei), and $\sigma^2 
\approx 0.015 A^{5/3} t$ is the spin cut-off parameter. The 
temperature $t$ is defined by $U=at^2-t$, where $a$ is the level density parameter. 

There exist no data on $a,\delta$ for $^{48}$Sc. If one adopts for $^{48}$Sc the values 
$a=5.96$ MeV$^{-1}$, $\delta=-2.37$ MeV obtained in Ref.~\cite{Dilg73} for $^{46}$Sc, one gets for the $J=0$ level density at $E_x=$6.8 MeV $\rho(0^+)+ \rho(0^-)\approx 59$ MeV$^{-1}$. However, this level density must be an overestimation, since at $E_x=$ 3 MeV the same parametrization gives $\rho(0^+)+ \rho(0^-)\approx 5$ MeV$^{-1}$, but not a single $J=0$ state is listed in the ENSDF database for $^{48}$Sc for $E_x<$3 MeV. A moderner parametrization $a=5.74$ MeV$^{-1}$, $\delta=-1.9$ MeV for $^{48}$Sc in the RIPL-2 database gives $\rho(0^+)+ \rho(0^-)\approx 33$ MeV$^{-1}$ at $E_x=$6.8 MeV (it still somehow overestimates the low-energy level density: $\rightarrow \rho(0^+)+ \rho(0^-)\approx 3$ MeV$^{-1}$ at $E_x=$ 3 MeV). Thus, one may realistically assume the mean level spacing of about 50--70 keV between the $0^+$ states of the normal isospin in the vicinity of the IAS in $^{48}$Sc.
Then, if the IAS were essentially spread over those $0^+$ states, the experiment~\cite{Gr07} would have been able to resolve components of the IAS fine structure. The fact that no fine structure was observed can easily be understood from a comparison of a typical $\Gamma^\downarrow_A$ of the order of few keV with a much larger mean level spacing.

To estimate the cross-section for the reaction $^{48}$Ti(n,p)$^{48}$Sc(IAS), we use in Eq.~(\ref{M}) the QRPA result $M^{0\nu}_F\approx 0.14$ fm$^{-1}$.
Then, the calculated ratio $\left|\frac{\langle IAS | \hat T^{+} | 0_f \rangle}{\langle IAS | \hat T^{-} | 0_i \rangle}\right|^2\approx 2\cdot 10^{-6}$, combined with the experimental cross-section $ \frac{d^2\sigma_{pn}}{d\Omega dE}\approx 10$ mb/(sr MeV) for the reaction $^{48}$Ca(p,n)$^{48}$Sc(IAS) at the proton energy $E_p=134$ MeV~\cite{And85}, leads to an estimate $\frac{d^2\sigma_{np}}{d\Omega dE}\approx 20$ nb/(sr MeV).
Note, that by choosing a smaller neutron incident energy this estimate can further be improved by a factor of 2--3, due to the increasing Fermi unit cross-section~\cite{tad87}.
The account for the Coulomb mixing of the IAS with the IVMR would modify the value of $\frac{d^2\sigma_{np}}{d\Omega dE}$ by few percents, as was estimated by making use of the method of Ref.~\cite{Rod10}, and therefore can be neglected. 

Note, that as a byproduct we can get a realistic estimate for the Fermi 
$2\nu\beta\beta$ NME for $^{48}$Ca: $M^{2\nu}_F\approx - \frac{e^2 M^{0\nu}_F}{2 \bar\omega_{IAS}^2}=-1.4\cdot 10^{-3}$ MeV$^{-1}$. Indeed, this value is much smaller than 
the Gamow-Teller NME $|M^{2\nu}_{GT}|\approx 0.05$ MeV$^{-1}$ governing the $\beta\beta$ decay of $^{48}$Ca~\cite{Bar10}.

To conclude, we have discussed here a possibility of a measurement of $M^{0\nu}_F$ 
for $^{48}$Ca  and estimated the cross-section of the appropriate reaction $^{48}$Ti(n,p)$^{48}$Sc(IAS).

The work is supported in part by the Deutsche Forschungsgemeinschaft.
The author gratefully acknowledges fruitful discussions of the relevant experimental issues with Profs.~H.~Ejiri, D.~Frekers, H.~Sakai, J.~Schiffer, and R.~Zegers.

\end{document}